\documentclass[twocolumn,showpacs,amsmath,amssymb,prl,aps,superscriptaddress]{revtex4-1}

\usepackage{graphicx}
\usepackage{dcolumn}
\usepackage{bm}

\usepackage{dcolumn}
\usepackage{bm}
\usepackage{hyperref}
\usepackage{mathtools}
\usepackage{appendix}
\usepackage{color}
\graphicspath{ {images/} }
\usepackage{braket}

\usepackage{braket}
\newcommand{\ee}{\end{equation}}
\newcommand{\be}{\begin{equation}}
\newcommand{\ea}{\end{eqnarray}}
\newcommand{\ba}{\begin{eqnarray}}
\newcommand{\ean}{\end{eqnarray*}}
\newcommand{\ban}{\begin{eqnarray*}}


\begin{document}

\title[Hydrodynamic superradiance]{Hydrodynamic superradiance in wave-mediated cooperative tunneling}
\thanks{We acknowledge the generous financial support of the NSF through grant CMMI-1727565 , the European Union’s Horizon 2020
Research and Innovation Program under the Marie Sklodowska-Curie project EnHydro, grant No. 841417, and the CNPq under (PQ-1B) 301949/2007-7 and FAPERJ Cientistas do Nosso Estado Project No. 150.036/2018. Authors tagged with  $^*$ have an equal contribution to this work.}

\author{K. Papatryfonos}

\affiliation{Gulliver UMR CNRS 7083, ESPCI Paris, Université PSL, 75005 Paris, France} 
\affiliation{Department of Mathematics, MIT, 77 Massachusetts Ave., Cambridge MA 02139, USA}

 \author{M. Ruelle$^*$}
 \affiliation{Gulliver UMR CNRS 7083, ESPCI Paris, Université PSL, 75005 Paris, France} 
  \author{C. Bourdiol$^*$}
  \affiliation{Gulliver UMR CNRS 7083, ESPCI Paris, Université PSL, 75005 Paris, France} 
\author{A. Nachbin}
 \email{nachbin@impa.br}
\affiliation{IMPA, Estrada Dona Castorina 110, Rio de Janeiro 22460-320, Brazil}

\author{J.W.M. Bush}
 \email{bush@math.mit.edu}
\affiliation{Department of Mathematics, MIT, 77 Massachusetts Ave., Cambridge MA 02139, USA}

\author{M. Labousse}
 \email{matthieu.labousse@espci.psl.eu}
\affiliation{Gulliver UMR CNRS 7083, ESPCI Paris, Université PSL, 75005 Paris, France}

\date{\today}
\begin{abstract}
Superradiance and subradiance occur in quantum optics when the emission rate of photons from multiple atoms is enhanced and diminished, respectively, owing to interaction between neighboring atoms.  We here demonstrate a classical analog thereof in a theoretical model of droplets walking on a vibrating bath.  Two droplets are confined to identical two-level systems, a pair of wells between which the drops may tunnel, joined by an intervening coupling cavity.  The resulting classical superradiance is rationalized in terms of the system's non-Markovian, pilot-wave dynamics.
\end{abstract}

\maketitle

\section{Introduction}
In classical physics, particle motion is characterized in terms of dynamical states which may provide in turn the basis for a statistical description of the system.
In quantum physics there is no such dynamical description: quantum systems are described entirely in terms of the evolution of their statistical states~\cite{Cohen-Tannoudji}. For a quantum system with at least two subsystems,
non-separability arises when the quantum state cannot be factored into a product of states of the individual subsystems~\cite{Popescu}.  Many well-known bipartite quantum phenomena arise as a result of such non-separable states. One is the phenomenon of superradiance and subradiance, whereby the probability of decay of two coupled systems, as marked by photon emission, depends on the distance between these systems \cite{Solano2017, DeVoe, Eschner2001,Dicke}.

Superradiance, a ‘cooperative’ spontaneous emission of photons from a collection of $N$ atoms, was theoretically predicted in 1954 by Robert Dicke \cite{Dicke}. When thermally excited atoms emit photons incoherently with respect to each other, the emitted intensity is proportional to the number of atoms, $N$. However, when the atoms radiate coherently, in phase with each other, the net electromagnetic field is proportional to $N$, and the emitted intensity thus scales as $N^2$. As a result, the atoms may decay at an enhanced rate that is up to $N$ times faster than for incoherent emission, a phenomenon termed `superradiance'. 

When the distance between the atoms is of the order of the emission wavelength, rationalizing this collective behavior requires consideration of the system's quantum nature. The most elementary demonstration of Dicke’s superradiance is achieved with two ions~\cite{DeVoe}. In this experiment, the photon emission rate $\Gamma(R)$ is characterized in terms of the inter-ion distance $R$, and compared to the single-ion emission rate $\Gamma_0$ in the same apparatus.
The quantum theory describes a bipartite system in which two interacting two-level systems form a single four-level system. When far apart, the two sub-systems are independent and can be either in the ground or excited states. When the two two-level systems interact, the collective system is treated as a single  non-factorable four-level system, consisting of a ground state $\ket{--}$, degenerate first excited states $\ket{+-}$ or $\ket{-+}$ that are energetically indistinguishable, and a second higher energy state $\ket{++}$. Here $\ket{+}$ and $\ket{-}$ denote the states of the two subsystems, nomenclature that we adopt for our system. The decay rates $\Gamma_{\pm}$ to or from the degenerate excited states $\ket{+-}$ and $\ket{-+}$ are well-approximated by $
\Gamma_\pm(R)= \Gamma_0 \left(1 \pm \frac{3}{2}\frac{\sin(kR)}{kR} + \ldots\right)$ in the limit $kR\gtrapprox 10$, 
where  $k=2\pi/\lambda$ denotes the wavenumber and $\lambda$ the emission wavelength ~\cite{DeVoe}. Specifically, $\Gamma_{+}$ and $\Gamma_{-}$ are the rates corresponding to transitions to or from the symmetric $(\ket{+-}$ + $\ket{-+})/\sqrt{2}$ and anti-symmetric $(\ket{+-}$ - $\ket{-+})/\sqrt{2}$ states, respectively. We note that $\Gamma_\pm$ can be either greater or less than the bare emission rate $\Gamma_0$, corresponding to, respectively, superradiance or  subradiance. The measured form of $\Gamma_{\pm}(R)$ was found to be in very good agreement with Dicke's theoretical prediction~\cite{DeVoe}. Such sinusoidally modulated, enhanced or diminished radiance is an indirect measure of the non-separability of states, a phenomenon commonly thought to be peculiar to the quantum realm. We here present a macroscopic hydrodynamic system that exhibits the defining features of superradiance.

We proceed by building an analog of superradiance by satisfying three criteria. First, we consider a classical two-level system in which one of the two states may be treated as the lower-energy state, in the sense that it is more likely to arise. Second, we produce a bipartite system, consisting of two such two-level systems, coupled in such a way that their collective behaviour cannot be specified in terms of a linear combination of its individual subsystems. Third, we demonstrate that the probability of transition from state to state in each subsystem varies in a sinusoidal fashion with distance between the two subsystems.

Hydrodynamics has long served as a rich source of physical analogs for electromagnetic and optical phenomena \cite{Young1804,Bragg1947,Bush2015b}.
More recently, hydrodynamic analogs of white~\cite{Whiteholes} and black holes~\cite{Unruh2002,Rousseaux} have been firmly established. Until recently, hydrodynamic analogs of quantum systems were relatively rare, but included the Aharanov-Bohm effect~\cite{Berry1980} and the Casimir effect ~\cite{Denardo2009}. The walking-droplet system discovered in 2005 by Yves Couder and Emmanuel Fort~\cite{Couder2005a,Couder2006} has proven to be a remarkably rich source of physical analogs for both optical and quantum systems. This pilot-wave hydrodynamic system exhibits many features previously thought to be exclusive to the quantum realm, and so has initiated the burgeoning field of hydrodynamic quantum analogs~\cite{Bush2015a,Bush2018Chaos,BushOza}. 

\begin{figure}
\includegraphics[width=0.5\textwidth]{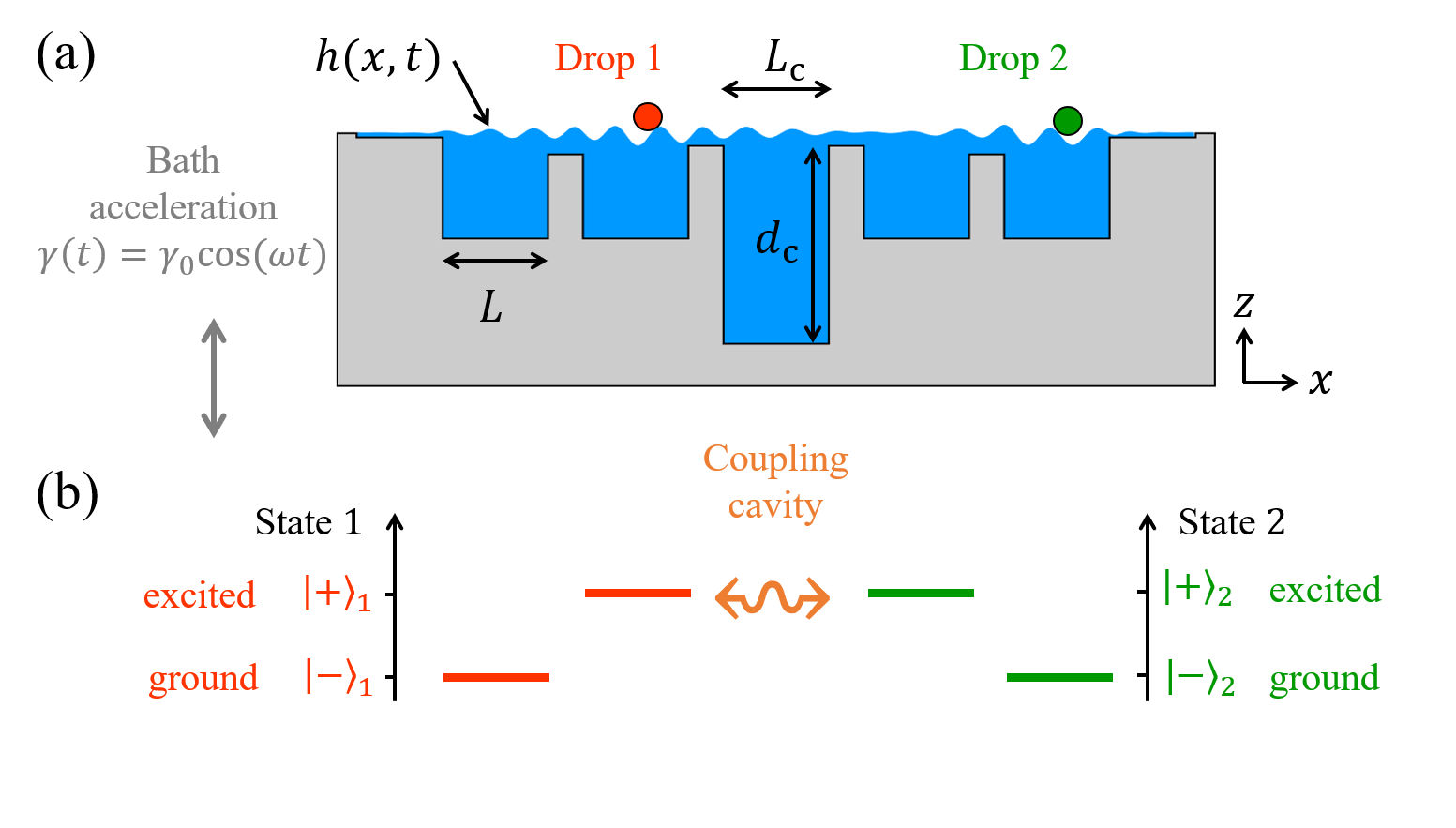}
\caption{{\bf Coupling two 2-levels systems}. (a) The system consists of a pair of droplets walking on the surface of a vibrating fluid bath. Each droplet is  confined to a pair of wells separated by barriers across which they may tunnel unpredictably. The extent of coupling in this bipartite tunneling system is prescribed by both the magnitude of the vertical vibrational forcing, $\gamma_0$, and the width, $L_c$, of the intervening coupling cavity. (b) The state of each droplet is denoted by $\ket{-}$ or $\ket{+}$ according to whether the drops are in, respectively, the outer `ground' state or the inner `excited' state.}
\label{fig:Fig1 NEWstyle}
\end{figure}

 Couder and Fort ~\cite{Couder2005a} discovered a classical pilot-wave system that consists of a millimetric droplet bouncing on the surface of a vibrating liquid bath, self-propelling by virtue of a resonant interaction with its own wave field. 
 By virtue of this resonance, the droplet is accompanied by a quasi-monochromatic wave field that imposes a dynamic constraint on the droplet that gives rises in the emergence of quantized dynamical states ~\cite{Bush2015a}. For example, quantized orbital states emerge when the walking droplets move in the presence of either Coriolis~\cite{Fort2010,Harris2014a,Oza2014b} or central spring forces~\cite{Perrard2014,Labousse2014,Labousse2016}. Moreover, the persistence of the pilot-wave field renders the system non-Markovian; specifically, the instantaneous wave force imparted to the drop during impact depends on the particle's history. The droplet thus navigates a potential landscape of its own making. This non-Markovian feature of the droplet dynamics gives rise to behavior that might be mistakenly inferred to be spatially non-local if the influence of the wave-field is not adequately resolved ~\cite{BushOza}  Of particular interest here is the hydrodynamic analog of unpredictable quantum  tunneling~\cite{Qdash}, as has been demonstrated both experimentally ~\cite{Eddi2009b,Tadrist2020} and numerically~\cite{Nachbin2017,Hubert2017}.

We explore here a bipartite system using the theoretical model developed by Nachbin {\it et al.} \cite{Nachbin2017} to describe tunneling in the hydrodynamic pilot-wave system. 
The two subsystems correspond to single particles confined to one of two cavities, the preferred cavity corresponding to the ground state, and the other to the excited state. The two subsystems are coupled by their mutual wavefield, which spans an intervening coupling cavity (Figure 1a). Transitions between states in the subsystems correspond to individual tunneling events, the rate of which depends on the distance between the two subsystems, specifically, the length of the coupling cavity, $L_c$. We proceed by demonstrating that the frequency of transitions oscillates with $L_c$, and so present a
classical analog of sub- and superradiance in quantum mechanics.

\section{Mathematical Model and Numerical Method} 
Nachbin et al. \cite{Nachbin2017} formulated a theoretical model for the one dimensional motion of a walking droplet over a vibrating liquid bath with complex topography. Here we adapt this model in order to consider two identical particles in the multiple-cavity geometry depicted schematically in figure \ref{fig:Fig1 NEWstyle}. The positions, $X_j$ ($j=1,2$), of the two identical particles of mass $m$ evolve according to Newton's Law:
\be
m\ddot{X}_j+ c~F(t)\dot{X}_j = - F(t)\; \frac{\partial\eta}{\partial x}(X_j(t),t).
\label{Drop1ODE}
\ee
The particle moves in response to gradients of the wave elevation $\eta(x,t)$, which thus plays the role of a time-dependent potential. The particle motion is resisted by a drag force proportional to its speed. The drag constant $c$ follows from the modeling presented in Molacek \& Bush \cite{Molacek2013b} and Milewksi {\it et al.}~\cite{Milewski2015}. The time dependence of these propulsive and drag forces is prescribed by $F(t)$, as arises in the walker system owing to the droplet's bouncing~\cite{Bush2015a,Milewski2015,Nachbin2017}. In terms of their lateral motion, the particles are viewed as oscillators that can transition unpredictably between two neighboring cavities. 

The particles serve as moving wave sources that establish their own time-dependent wave potential that is computed as follows. The velocity potential of the liquid bath $\phi(x,z,t)$ is a harmonic function, i.e. satisfying Laplace's equation. In the bulk of the fluid, the velocity field is given by $(u,v)=\nabla\phi$. The fluid bath of density 0.95 g/cm$^3$, viscosity 16cS and surface tension 20.9 dynes/cm oscillates vertically with amplitude $A_0$ and frequency $\omega_0 = 80$Hz. The resonant bouncing of the particle at the Faraday frequency triggers a monochromatic damped wave pattern with a corresponding deep-water wavelength of $\lambda_F=4.75$~mm. 
The wave model is formulated in the bath's reference frame, where the effective gravity is $g(t)= g(1+\gamma_0 \sin(\omega_0 t))$, where $g$ is the acceleration due to gravity, and $\gamma_0 = A_0 \omega_0^2/(4\pi^2$) is the maximum vibrational acceleration. The wave field thus evolves according to \cite{Milewski2015,Nachbin2017}: 
\be
\frac{\partial \eta}{\partial t} = \frac{\partial \phi}{\partial z} + 2\nu 
\frac{\partial^2\eta}{\partial x^2},
\label{Kin}
\ee
\be
\frac{\partial \phi}{\partial t}  =  - g(t) \eta + \frac{\sigma}{\rho} \frac{\partial^2\eta}{\partial x^2} + 2\nu 
\frac{\partial^2\phi}{\partial x^2} -\sum_{j=1,2} \frac{P_d(x-X_j(t))}{\rho}
\label{Bern}
\ee
The particles ($j=1,2$) generate waves on the free surface by applying local pressure terms $P_d$. The wave forcing term $P_d(x-X_j(t))$ and the coefficient $F(t)$ are activated 
only during a fraction of the Faraday period $T_F$, corresponding to the contact time $T_c$ in the walking-droplet system and approximated by $T_c=T_F/4$. The particle is assumed to be in resonance with the
most unstable (subharmonic) Faraday mode of the bath~\cite{Milewski2015}, a key feature of pilot-wave hydrodynamics \cite{Protiere2006,Bush2015a,BushOza}.

By adjusting the system geometry, specifically the barrier widths and depths and the aspect ratios of the cavities, we are able to design a bipartite system consisting of two coupled two-cavity subsystems. One particle is confined to each of these two subsystems, but may tunnel between the inner and outer cavities within it. The likelihood of finding the particles in the inner or outer cavities depends on the relative magnitudes of the system energy when the particles are in these cavities. Each subsystem displays metastability, in the sense that the particle may find itself in either the energetically unfavorable, less stable `excited' state corresponding to one cavity, or the energetically favorable, more stable `ground' state corresponding to the other. 

In equations (\ref{Drop1ODE}) and (\ref{Kin})-(\ref{Bern}), 
spatial derivatives are computed using the Fast Fourier Transform (FFT) in $x$. The lateral shallow parts of the fluid domain extend 
sufficiently far to ensure periodic (quiescent) conditions in the far field. The vertical speed at the free surface $\phi_z(x,0,t)$ is defined through a Dirichlet-to-Neumann (DtN) operator~\cite{Nachbin2017} and yields a Fourier integral operator computed in a straightforward fashion through an FFT. The DtN operator mathematically reduces the two-dimensional fluid problem to one spatial variable defined along the undisturbed free surface. To compute the DtN operator, a numerical conformal mapping is performed that maps the $(x,z)$ fluid domain onto a ($\xi,\zeta$) canonical flat strip. Details of the conformal mapping can be found in  \cite{SIAP01,FokasNachbin}.
The mapping  is computed only once and provides the relation 
$x=x(\xi)$ along the undisturbed free surface.
We denote by $\cal{F}$ the FFT in the $\xi$-coordinate, which runs along the undisturbed free surface in the canonical domain. We denote by $\phi(x,0,t)) = \varphi(x,t)$ the Dirichlet data. We thus have that
\be
\phi_z(x,0,t) = DtN[\phi](x,t)=\frac{{\cal{F}}^{-1} [G(k){\cal{F}}[\varphi(\xi)]]}{M(\xi(x,0))},
\label{DtNFFT}
\ee
where $G(k)=k\tanh{k}$ is the Fourier multiplier~\cite{Nachbin2017}. The metric coefficient is $M(\xi) = \sqrt{|J|}$, where
$|J|$ is the Jacobian of the (($x,z)\rightarrow (\xi,\zeta$)) change of variables, evaluated along the undisturbed free surface~\cite{SIAP01,FokasNachbin}.
 To summarize, the geometrical information of the cavities and barriers is  encoded in $M$ 
and in $\xi=\xi(x,0)$, which is obtained with the inverse map. The time evolution is performed with a 
second-order fractional-step Verlet method~\cite{Nachbin2017}. To confirm the statistical significance of our results, we performed 6 runs with durations of 48000 Faraday periods for each geometrical configuration. We used random initial conditions for the particle positions, and discarded the first 10\% of the runs in order to eliminate any trace of transient effects.

We proceed by examining the system energetics.  Energy is transferred in a complex fashion between particle and wave. In certain regimes, where either periodic or statistically steady dynamics emerge, the wave energy input from the bath vibration balances the dissipative losses associated with the fluid viscosity. We here characterize the particle energy in phase space, in instances where nearly periodic cycles are observed. The excited state will be marked by relatively large particle excursions within the cavities. We define a cycle-averaged energy quantity for particle $j=1,2$
 \be
\bar{A}_j = \int_t^{t+T} \left [ \Omega_0 X_j^2(t) + \Omega_0^{-1} \dot{X}_j^2(t) \right ] dt 
\ee
that we refer to as the `oscillator action', a name originating from conserved quantities used to describe wave propagation in fluids~\cite{Andrews78}. We note that the dimension of $\bar{A}^{1/2}$, a length, is related to the range of droplet excursions within a cavity. Hence it indicates whether the cavity is an excited state. 
 
The oscillator action has both kinetic energy and potential energy components.
Recall that for a particle of mass $m$ in a harmonic oscillator with natural frequency $\Omega_0$, $\Omega_0 X_j^2(t) + \Omega_0^{-1} \dot{X}_j^2 $ is constant and defines an elliptical trajectory in phase space. The constant is a multiple of the oscillator action $A=H/2 m \Omega_0$, where 
$H=\frac{1}{2}( m\dot{X}^2+m\Omega_0^2 X^2)$ is the harmonic oscillator's Hamiltonian. For our driven dissipative system, we have neither a Hamiltonian  nor a constant action over each cycle. Numerically we 
compute
\be
\bar{A}_j = \sum_n  \left [ \bar{\Omega}_0 X_j^2(t_n) + \bar{\Omega}_0^{-1} \dot{X}_j^2(t_n) \right ] \Delta T,
\ee
where $\Delta T = T_F$, so the discrete step corresponds to the Faraday period. This diagnostic, $\bar{A}$, is calculated over each cycle, whose approximate frequency $\bar{\Omega}_0$ is readily computed numerically. The energetic barrier to tunneling between two states is denoted $\delta A_{\mathrm{total}}$ (see Supplemental Materials). A detailed analysis of this diagnostic reveals that the oscillator action ($\bar{A}$) remains approximately constant in each cavity for both the single and two-droplet cases, for each specific geometry considered here. In addition, $\bar{A}$ is always lower in the ground state than in the excited state in all geometries considered here. These features ensure that $\bar{A}$ is a reliable diagnostic to quantitatively characterize the energetics in each subsystem. Figure S1 of the Supplementary Materials (SM) details the behaviour of the kinetic and potential components of $\bar{A}$ as a function of the coupling cavity length for the single-drop case, whereas Figure S2 of the SM analyses in detail the two-drop case. 

Another diagnostic will be 
the underlying wave intensity, which will be monitored as the particle moves between its ground and excited states. Given spatial grid-points, $x_k$  ($k=1,\ldots ,K$), the wave intensity at time $t_n$ is defined by
\begin{equation}
\vert\vert\eta\vert\vert (t_n) =  \left( \sum_{k=1}^K \left(\eta(x_k,t_n)\right)^2  \right)^{1/2}.
\label{wave_intensity}
\end{equation}

\section{Results}
Our system comprises two subsystems, labelled 1 and 2 in Figure~\ref{fig:Fig1 NEWstyle}a, each consisting of a particle in an identical pair of cavities separated by a barrier across which the particle may tunnel. The two subsystems are separated by a coupling cavity of variable width $L_c$, and by barriers that are sufficiently high as to preclude the particles from tunneling into the coupling cavity. The particles are thus confined to one of the two subsystems, but may find themselves in either the inner or outer cavities of their subsystem. Each of the four cavities has a fixed length of 1.2 cm, corresponding to approximately 2.5$\lambda_F$. The waves are transmitted across the central cavity, and so provide the coupling between subsystems 1 and 2. The efficiency of this coupling is prescribed by the geometry of the central cavity: by increasing its depth $d_c$, the coupling is increased, allowing the coupling cavity to serve as a nearly resonant transmission line \cite{nachbin2018}. In all simulations, we thus set the coupling cavity depth to $d_c=6.3~\lambda_F $ which ensures strong inter-cavity coupling. In subsystems 1 and 2, the particle may tunnel between the outer and inner cavities, but the probability of presence in these cavities is not generally 50\%. Specifically, in the geometry considered, the particles are more likely to be in the outer cavity. 

We may thus describe this bipartite tunneling system in terms of two coupled, two-level systems, as shown in Fig.~\ref{fig:Fig1 NEWstyle}b. 
It is important to note that the energetic structure in our system is tunable and may be tailored in a predictable fashion. By simply adjusting the geometry, the stability of the cavities may be altered and even reversed. In particular, as detailed in Figure S3 of the SM, by changing the length of the four outer cavities it is possible to choose which of the inner or the outer cavities will be energetically favourable and thus act as the ground state. The magnitude of their energy difference may thus be adjusted and even eliminated, in which case the particles can be found in the inner and outer cavities with equal probability. 
\begin{figure}
\centering
\includegraphics[width =0.5\textwidth]{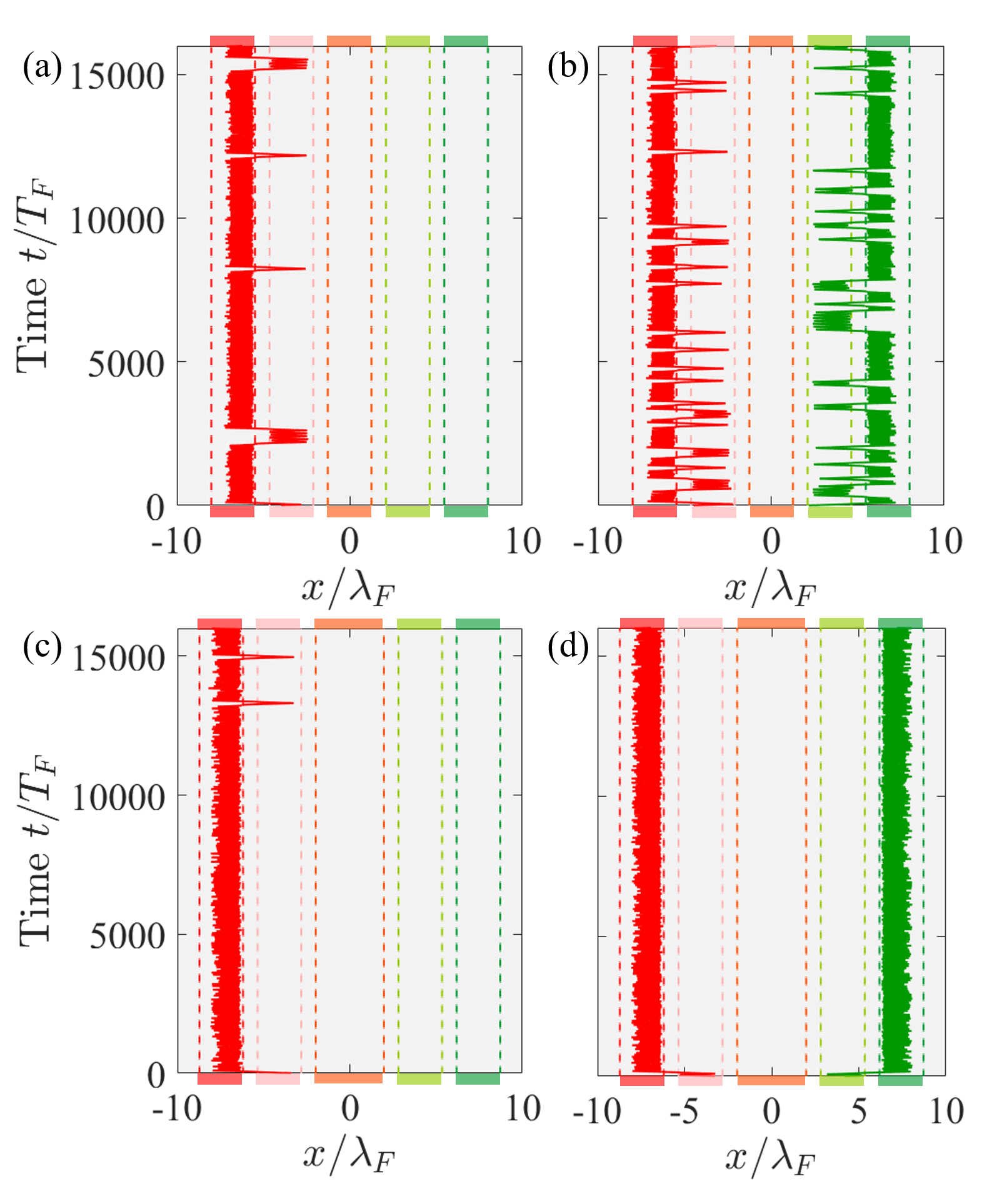}
\caption{(a) A single droplet tunnels between its ground state (outer cavity) and excited state (inner cavity). The middle cavity (marked in orange) couples the left and right subsystems. (b) In the bipartite system, a pair of drops tunnel between their ground and excited states. The red and green lines denote the particle paths in the left and right subsystems, respectively. In (a) and (b) the coupling cavity length is $L_c=2.44~\lambda_F$ and the four other cavities lengths are $L=2.52~\lambda_F$. Comparing (a) and (b) makes clear that a single particle's tunneling probability is substantially increased by the presence of a neighboring drop, corresponding to superradiance. The configuration shown in (c) and (d) is similar to (a) and (b) but with a larger coupling cavity $L_c=3.96~\lambda_F$. Comparing (c) and (d) reveals that the particle's tunneling between cavities is inhibited by the presence of the neighboring drop, an effect corresponding to subradiance.}
\label{fig:Fig2}
\end{figure}

\begin{figure*}
\centering
\includegraphics[width =\textwidth]{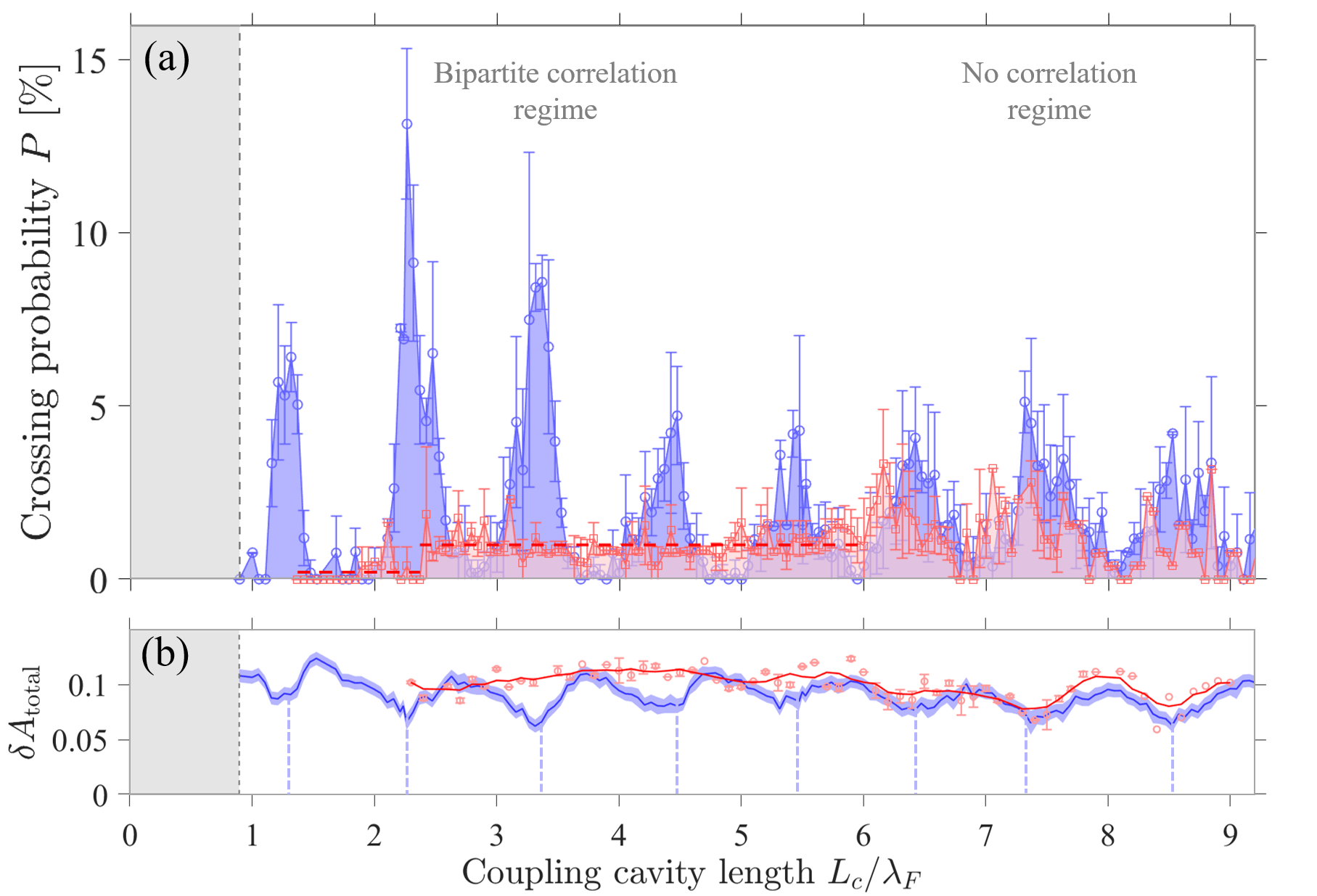}
\caption{(a) The dependence of the probability of tunneling from the outer to the inner cavity on the width of the coupling cavity, $L_c$, for a single drop (red $\square$) and
for two drops (blue $\circ$). Red dashed line is the average probability for a single drop over cavity lengths such that $L_c/\lambda_F < 6$, specifically, $\langle P\rangle=1.0\%$ ($95\%$ confidence interval $[0.91:1.09]$). (b) Evolution of the energy barrier to tunneling, $\delta A_{\mathrm{total}}$, for the two-droplet system (blue) and the one-droplet system (red). Blue shaded regions adjoining the lines denote the error bars. Note that the probability maxima in panel (a) correspond to energy-barrier minima in (b). For $L_c/\lambda_F > 6$, the tunneling probability of a particular drop converges to that of a single droplet, and so is evidently unaffected by its distant neighbor. The grey region on the left denotes the region of tall, slender coupling cavities that could not be reliably explored numerically.}
\label{fig:Fig3}
\end{figure*}

When a single droplet is placed in one of the two subsystems, the single-particle tunneling probability ($\Gamma_0$) may be deduced by counting the number of tunnelling events per attempt over a sufficiently long time interval. When droplets are placed in each of the two subsystems, the tunnelling probabilities change substantially from $\Gamma_0$, owing to the wave-mediated coupling of the two subsystems. Figure 2ab indicates that the probability of tunnelling increases substantially when the second droplet is present in the neighbouring system, an effect akin to superradiance. Figure 2cd presents a configuration with a longer coupling cavity in which the tunnelling probability is substantially decreased by the addition of a neighbor, corresponding to subradiance.

Having established that wave-mediated coupling alters the tunneling probability in this bipartite system, we proceed by characterizing the dependence of the transition probabilities on the system geometry. Figure 3a illustrates how the tunnelling probability depends on the length of the coupling cavity, $L_c$. The tunneling probability either increases or decreases relative to $\Gamma_0$, varying continuously with $L_c$ in an oscillatory fashion, reminiscent of the modulation of the emission rate reported in quantum mechanics superradiance~\cite{DeVoe}. Further to this modulation, the peaks become less intense and slightly broader as $L_c$ increases, a trend that is likewise apparent in QM superradiance~\cite{DeVoe}. 
For $L_c/\lambda_F > 6$, the single-drop and droplet-pair results converge, indicating that the two components of this bipartite system then effectively uncouple. Specifically, the tunneling probability of a particular drop is determined by the length of the coupling cavity, but unaffected by its distant neighbor.

The evolution of the transition probability with coupling cavity length is rationalized in Fig.~\ref{fig:Fig3}b, where we report the difference in oscillator action $\delta A_{\mathrm{total}}$ between the excited and ground states. A comparison between Figures~\ref{fig:Fig3}a and ~\ref{fig:Fig3}b shows a direct correspondence between the maxima and minima of the transition probability and the minima and maxima of $\delta A_{\mathrm{total}}$, respectively. We conclude that the wave-mediated lowering of the oscillator action strongly favors the transition from the ground to the excited state. The one-droplet system simulations reveal a difference of oscillator action $\delta A_{\mathrm{total}}$ that is approximately constant for sufficiently small coupling cavity length $L_c<6 \lambda_F$. 

We further note that when $L_c/\lambda_F <1$ the mid-cavity is a relatively deep, slender region, which 
poses problems for the nonlinear solver of the numerical conformal mapping. This limitation is presumably due to the crowding phenomenon, a numerical conformal mapping feature arising when two pre-images of 
vertices get exponentially close in the canonical domain \cite{SCh}. 
The gray region in Figure 3 identifies these slender coupling cavities that could not be reliably explored numerically. 

\begin{figure*}
\centering
\includegraphics[width
=\textwidth]{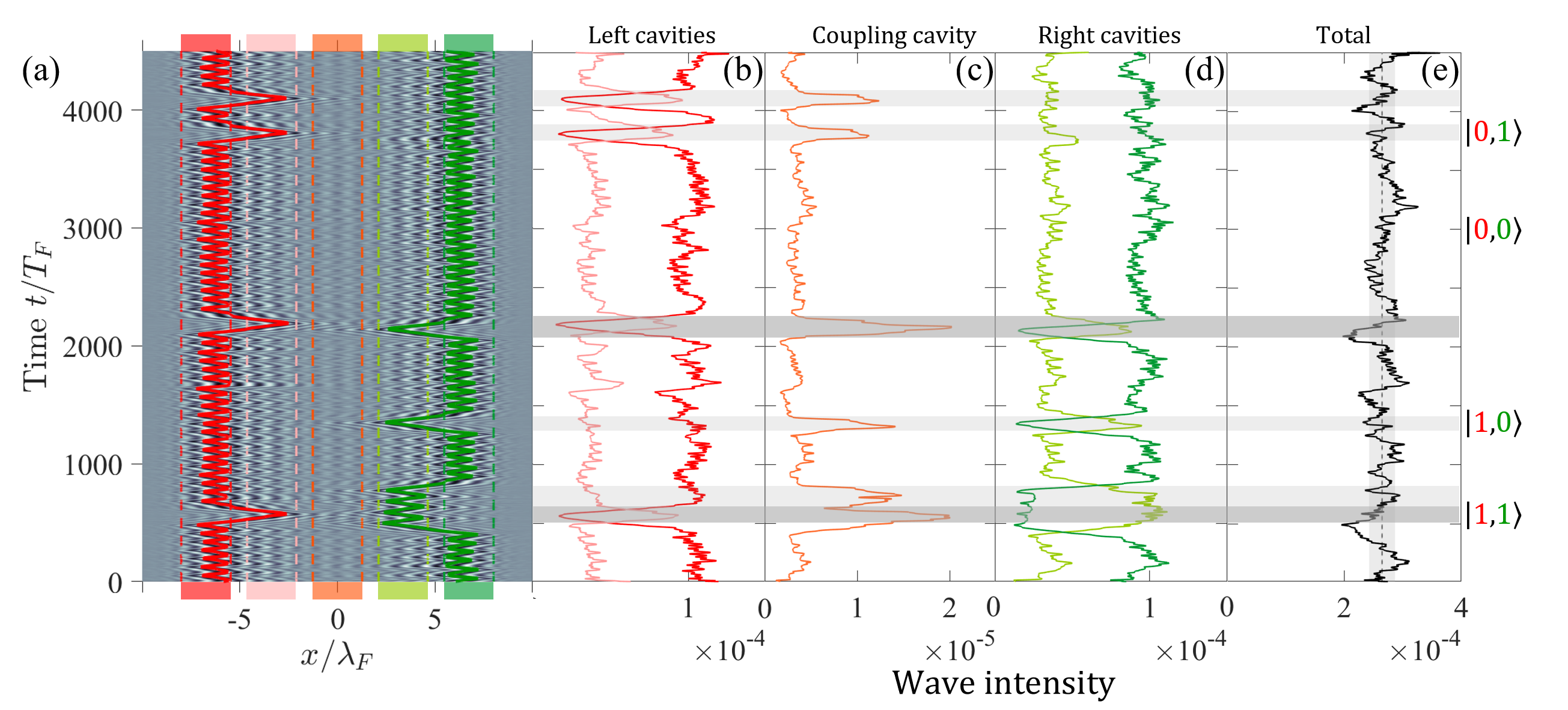}
\caption{(a) The particle paths and associated wave fields. White and black correspond to wave minima and maxima, respectively. (b)-(e) The associated wave intensity, as computed from Eq.~\ref{wave_intensity}. Comparison of (a) and (b) reveals the correlation between extrema in the wave intensity and the tunneling events. In (b)-(e), intervals in which one or both drops are in an excited state are denoted by light and dark grey, respectively, and are correlated with local extrema in the total wave intensity.}
\label{fig:Zoomed_in_wave_dashed}
\end{figure*}

Finally, we examine the correlation between the wave field structure and the particle dynamics. Figure 4a illustrates both the particle positions and instantaneous wave field. An assessment of the wave intensity (Eq.~\ref{wave_intensity}) in each cavity is shown in Figure 4(b-d). Each cavity is color-coded for ease of reference. The rightmost panel shows the wave intensity of the 
system across all cavities. Periods when one particle has tunneled into an excited state is denoted by light grey.
Such intervals are accompanied by a reversal in the relative wave intensity in the inner and outer cavities, 
and an increase in the wave intensity in the coupling cavity. Periods when both particles are in excited states are 
denoted by dark grey, and marked by maxima in the wave intensity in the coupling cavity and minima in the global wave intensity. The wave intensity diagnostic thus provides rationale for associating the outer and inner cavities with, respectively, ground and excited states.

\section{ Discussion } 
We have examined a model of bipartite tunneling in a classical system. The tunneling probability within the two subsystems is mediated by their common wave field and shown to depend on the length of the coupling cavity. This dependence is similar to that of the probability of photon emission on inter-ion spacing in the case of quantum superradiance ({\it e.g.} compare our Figure 2 with Figures 2 and 6 of Ref. \cite{DeVoe}). 
If we consider that the energy stored in our system is altered each time the droplet tunnels between the ground and excited states, we can establish a direct analogy between our system and quantum superradiance. Specifically, we may identify the alteration of the wave field that accompanies the transition from a high energy to lower energy state with the emission of a photon in QM. Thus, the probability of tunnelling is directly analogous to the transition rate in superradiant photon emission.

We have highlighted the similarities between our classical system of bipartite tunneling and super- or subradiant photon emission from ion pairs in magnetic traps. To account for the optical superradiance in QM, one must invoke the concept of collective, non-separable, states that are formed between the two ions, which are viewed as absorbing or emitting each photon collectively~\cite{scully, DeVoe, makarov_spontaneous_2003,makarov_metastable_2004}. QM offers no physical picture for the origins of non-separable states, only the mathematical tools required to calculate the relevant probability densities. Our system illustrates how such correlations may arise in a classical, wave-mediated system. The correlations in our system arise from the wave coupling between the two tunneling subsystems. While the form of the waves may be deduced by linear superposition, their influence on tunneling probability is most certainly non-linear. Specifically, the discrete tunneling events are unpredictable, as they depend in a subtle fashion on the interactions between the particles, the waves and the underlying potential wells formed by bottom topography~\cite{Eddi2008,Nachbin2017,Tadrist2020}. 

This bipartite tunneling system has given rise to collective
transition statistics, specifically 
subradiant and superradiant tunneling rates between the ground and excited states, which may be predictably controlled by altering the distance between the two two-level systems. Notably, the precise energetic difference between the ground and excited states may be altered and even reversed by changing the geometry of the outer cavities (Fig. S3). Our system thus provides a new platform for characterizing   wave-induced correlations between particle pairs. In particular, by taking the system geometry as a proxy for measurement settings, we may test the viability of violating  Bell-type inequalities with this classical pilot-wave  system~\cite{Vervoort2018}. 

\bibliographystyle{unsrt}
\bibliography{ARFMBib}

\subsection*{Supplemental materials: Analysis of the particle energetics}
In order to characterize the dependence of the system energetics on the underlying topography, we compute the oscillator action $\bar{A}$ (Eq.6). Specifically, for each particle, $\bar{A}$ was calculated for each oscillator cycle in phase space, for both the ground and excited states. 

We first studied the variations of $\bar{A}$ with the coupling cavity length $L_c$ when the outer cavity length was held fixed, which allowed us to correlate these variations with those in tunneling probability (see Figure 3). The results for the single droplet and droplet pairs are shown, respectively, in Figures S1 and S2. In the single-droplet case (Figure S1), the potential component of the oscillator action is constant, while the kinetic component of the oscillator action (and thus the total action) is always lower in the ground state (the outer cavity) than in the excited state (the inner cavity). This dependence establishes the oscillator action as a robust diagnostic for the energetics of our system. Notably, there are geometries in which the ground and excited states are reversed (see Figure S3), specifically the inner cavities correspond to the ground states. Here again, the action is a robust diagnostic, being consistently lower in the ground state than the excited state.

We proceed by detailing the oscillator action for the two-droplet case. As shown in Figure S2, when a second droplet is introduced in the system, the difference in oscillator action $\delta A_{\mathrm{total}}$ between the excited and ground states oscillates with the coupling cavity length, $L_c$. The oscillator action in the inner cavity is constant, independent of $L_c$, and we take its mean value $A^-$ as a reference. Conversely, the oscillator action in the outer cavity varies significantly as a function of the coupling cavity length $L_c$. We denote its value by $A^+(L_c)$. The action difference is expressed as
\begin{equation}
    \delta A_{\mathrm{total}}(L_c)=A^+-A^-.
    \label{Action_definition}
\end{equation}

Fig. S2 shows that the fluctuations of $\delta A_{\mathrm{total}}$ with $L_c$ are due primarily to those in the potential component. Maxima in the oscillations of $\delta A_{\mathrm{total}}$ coincide with minima in the tunneling probability (see Fig. 3a). Thus, as expected, when the difference in action, and so the energy difference between the two states, is higher, there is a lower probability of tunneling into the excited states (see Fig. 3b). 

\vspace*{0.1in}

\subsection*{Supplemental materials: Exploring other regimes by varying the topography. }
Varying the lengths of the four outer cavities also changes the tunneling probabilities. Indeed, one may even reverse the relative energies of the inner and outer cavities, changing the outer and inner cavities from ground to excited states (see Fig. S3a\textcolor{blue}{, c,} and e). Intermediate regimes also exist, wherein all cavities are equally likely to contain a drop, in which case the energy levels of the two subsystems become degenerate (see Fig. S3b and d).

\begin{figure*}
\includegraphics[width=\textwidth]{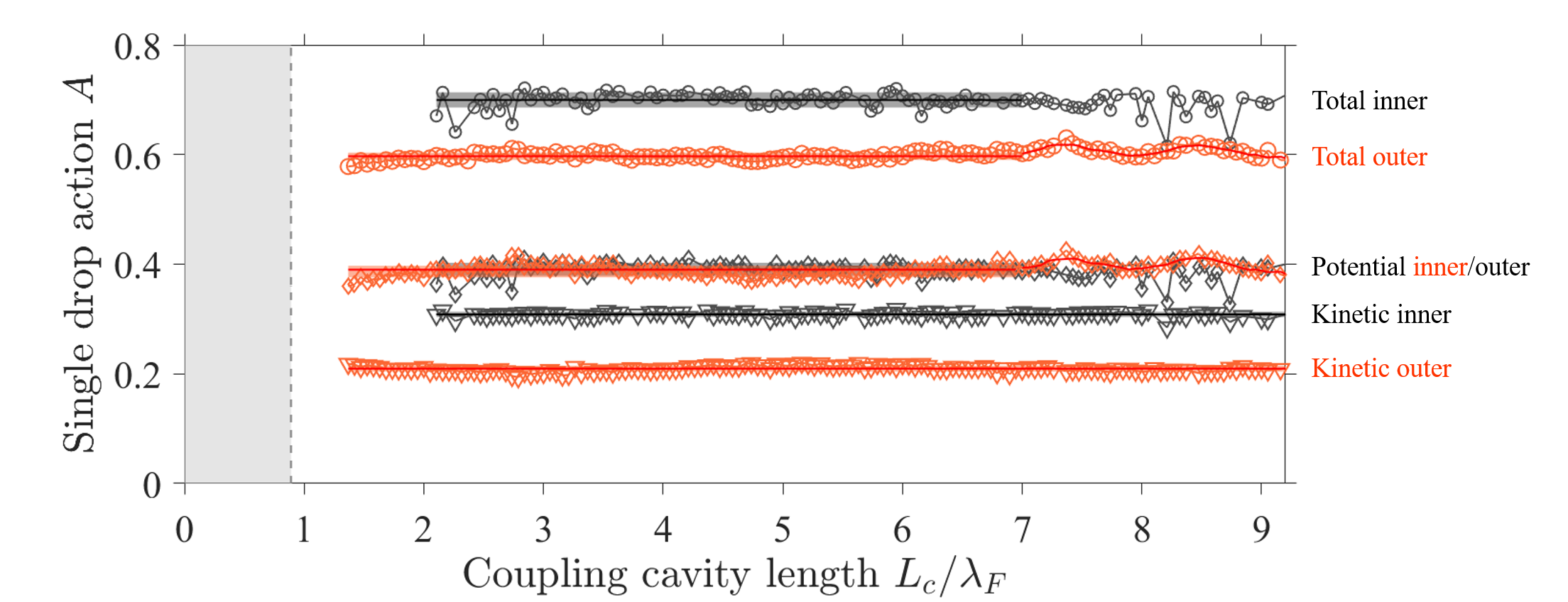}
\caption{Supplemental Figure 1: {\bf Oscillator action levels in the single drop system}. Neither the kinetic nor the potential components of the oscillator action change with the coupling cavity length, $L_c$. The potential energy is the same in the ground and excited states. However, the kinetic component (and thus the total action) is  always lower in the outer ground state than the inner excited state for the geometries considered. The length of the outer cavities is fixed at 2.53$\lambda_F$, their depth at 1.05$\lambda_F$.}
\label{fig:SMFig1}
\end{figure*}

\begin{figure*}
\includegraphics[width=\textwidth]{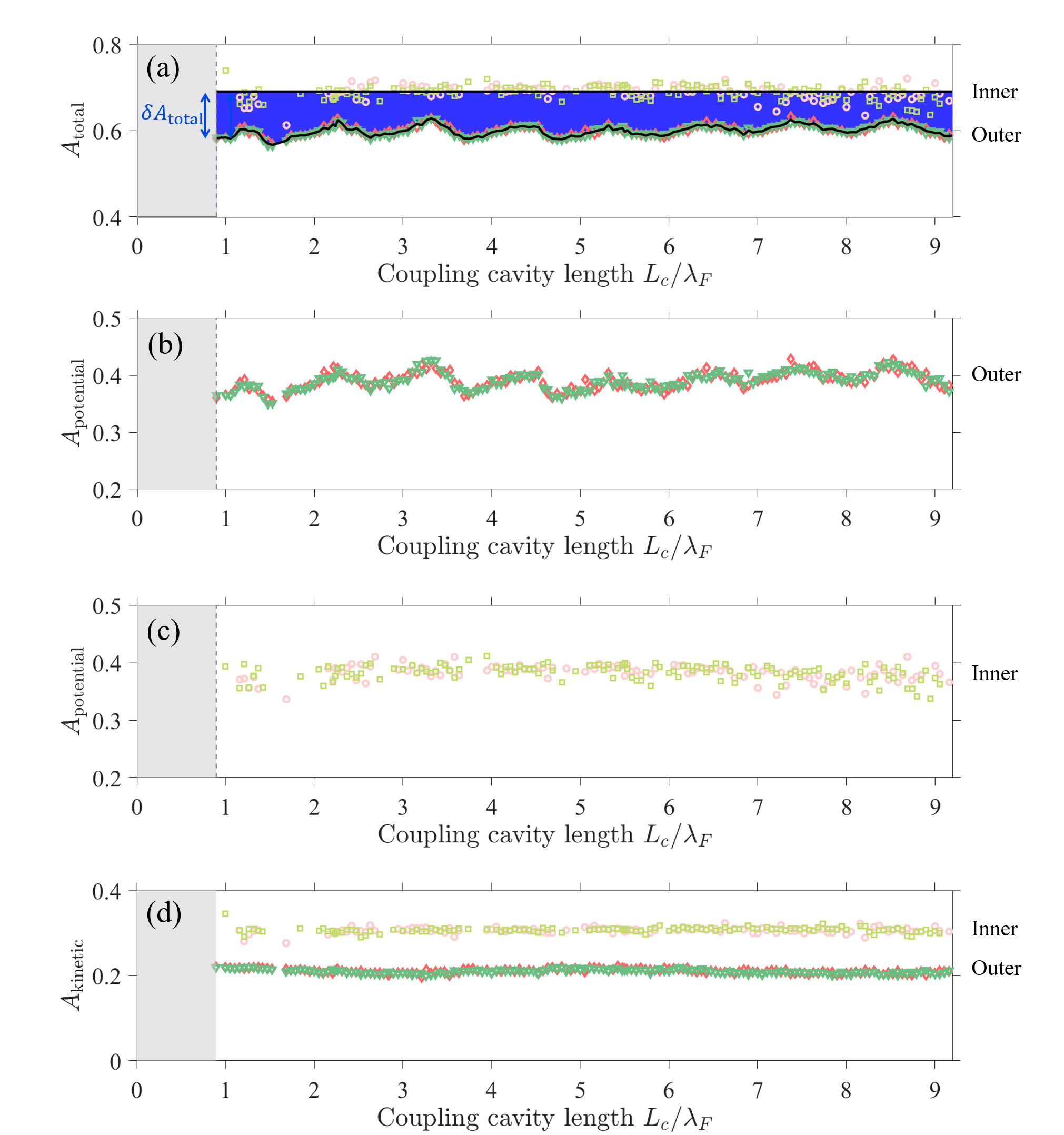}
\caption{Supplemental Figure 2: {\bf Oscillator action levels in the two-drop system}. With a second droplet in the system, the difference in oscillator action, $\delta A_{\mathrm{total}}$ (Eq.~\ref{Action_definition}), between the ground and excited states oscillates as a function of the coupling cavity length, $L_c$. The maxima in the oscillations of $\delta A_{\mathrm{total}} (L_c)$ coincide with the minima in the tunneling probability shown in Fig. 3(a) of the main text: when the energy difference between the two states is higher, there is a lower probability of tunneling into the excited states. (a) Action levels in the outer (ground) and inner (excited) states as a function of $L_c$. The blue area highlights their difference, $\delta A_{\mathrm{total}}(L_c)$. The dependence on $L_c$ of the potential part of the oscillator action in the (b) outer and (c) inner cavity. (d) The dependence on $L_c$ of the kinetic part of the oscillator action in the outer and inner cavities. Color-coding (see Fig. 1):  The red and pink symbols in (a)-(d) correspond to the drop on the left, the green and light-green symbols to the drop on the right. }
\label{fig:SMFig2}
\end{figure*}

\begin{figure*}
\includegraphics[width=\textwidth]{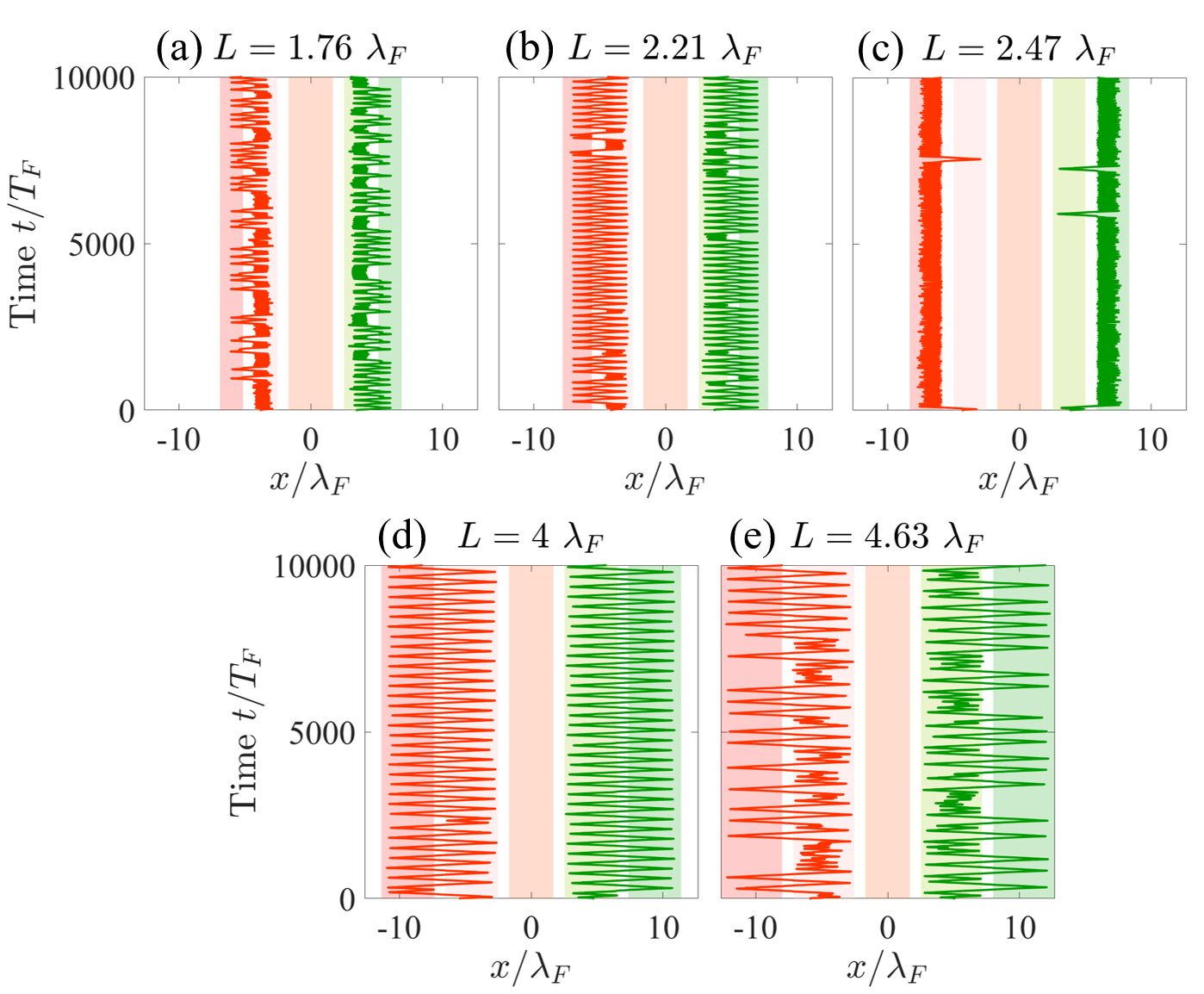}
\caption{Supplemental Figure 3: {\bf Energy level inversion by altering the length of the four outer cavities.}
The length of the coupling cavity was fixed at $L_c$=3.37$\lambda_F$ while that of the outer cavities, $L$, was varied: $L/\lambda_F=$
(a)1.75, (b) 2.21, (c) 2.47, (d) 4.0 and (e) 4.63. Changing $L$ can reverse the relative stability of the cavities. Specifically, as  $L$ increases progressively, the droplet's preferred cavity changes from the inner in (a) to the outer in (c), then again to the inner in (e). In the intervening geometries, (b) and (d), the inner and outer cavities are equally probable.}

\label{fig:SMFig3}
\end{figure*}

\end{document}